\documentclass[12pt,preprint]{aastex} 
\shorttitle{??}
\shortauthors{Reale}
\usepackage{amsmath}            

\begin{document}
\def\ntilde{\hbox{\rm n}}
\def\vv{\hbox{\bf v}}
\def\gvec{\hbox{\bf g }}
\def\rvec{\hbox{\bf r }}
\def\svec{\hbox{\bf s }}
\def\vvec{\hbox{\bf v }}

\title{More on the determination of the coronal heating function 
from Yohkoh data}
\author{F. Reale}
\affil{Dipartimento di Scienze Fisiche \& Astronomiche, Sezione di
Astronomia, Universit\`a di Palermo, Piazza del Parlamento 1, 
I-90134 Palermo, Italy \\ \email{reale@astropa.unipa.it}
}


\begin{abstract}
Two recent works have analyzed a solar large and steady coronal loop
observed with Yohkoh/SXT in two filter passbands to infer the
distribution of the heating along it. Priest et al. (2000) modelled the
distribution of the temperature obtained from filter ratio method with
an analytical approach, and concluded that the heating was uniform
along the loop. Aschwanden (2001) found that a uniform heating led to
an unreasonably large plasma column depth along the line of sight, and,
using a two component loop model, that a footpoint-heated model loop
(with a minor cool component) yields more acceptable physical
solutions.  We revisit the analysis of the same loop system,
considering conventional hydrostatic single loop models with uniformly
distributed heating, and with heating localized at the footpoints and
at the apex, and an unstructured background contribution extrapolated
from the region below the analyzed loop.  The flux profiles synthesized
from the loop models have been compared in detail with those observed
in both filter passbands with and without background subtraction; we
find that background-subtracted data are fitted with acceptable
statistical significance by a model of relatively hot loop ($\sim 3.7$
MK) heated at the apex, with a column depth $\sim 1/10$ of the loop
length. In discussing our results, we put warnings on the importance of
aspects of data analysis and modeling, such as considering diffuse
background emission in complex loop regions.

\end{abstract}



\section{Introduction}
\label{sec:intro}

The most recent solar X-ray and UV missions (Yohkoh, SoHO, TRACE) have 
produced a significant amount of spectacular data and images of the 
corona. The improved spatial resolution and sensitivity of the detectors 
have revealed the highly structured and variable nature of the
corona and have stimulated the interest of solar physics community. 

Whereas the morphological interpretation of the images is often
straightforward and has highlighted the presence of a variety of new
structures and phenomena, the physical interpretation of the data and
the inference of indications about items such as the heating of the
corona is a much more delicate matter. A possible problem is that many
physical and instrumental effects act concurrently and non-linearly.
The coronal plasma is highly thermally conductive {\it along}
the magnetic field lines but much less {\it across} them.  As a result,
non-uniform and localized heating sources which ignite coronal loops
will at the same time be elusive, because the temperature increases
rapidly along the whole loop, and determine a significant thermal
stratification across the loop and along the line of sight. Since the
coronal plasma is mostly optically thin, a telescope observing a
loop will then collect the emission from multiple thermal components
along the line of sight, filter them with different weights, depending
on its passband, and sum them up to yield a single data number for each
image pixel. The reconstruction of the original thermal structure of
the plasma column in a pixel is a difficult task and some attempts have
been made possible by multi-line observations with SoHO (Schmelz et
al.  2001). When analyzing data from wide-band imaging instruments such
as Yohkoh/SXT, one approach is to obtain maps of the weighted average
temperature of the various thermal components along the line of sight.
The interpretation of temperature distributions is somewhat easier
along bright loop structures, because there one has the reasonable
expectation that one thermal plasma component has a much larger
emission measure than the others, and therefore dominates the emission
and can be determined reliably.

This vision has driven recent studies of the thermal structure along
coronal loops, and of its usage as indicator of the distribution of the
heating deposition within the loop (Priest et al. 2000, Aschwanden
2001). The concept of these works was to compare the temperature
distributions of bright coronal loops derived from the data with those
obtained from models of thermally structured single coronal loops (or
combinations of them).  Priest et al. (2000) used an analytical approach
and devised a method to infer the heating distribution along the loops
from the analysis of temperature profiles, showing that along several
bright steady-state large-scale loops observed at the solar limb the
temperature distribution indicates uniformly distributed heating.
Aschwanden (2001) revisited the analysis of one of the loops selected
by Priest et al. (2000), and found that the previous approach led to
inconsistent results and, in particular, to an unreasonably large
plasma column depth required to match the emission measure inferred
from the data.  Using a two component loop model to explain the
observations, he also found that the data are consistently fitted
instead by a model coronal loop with a heating highly localized at the
footpoints and with a cool background component.

In this work we present a further analysis of the same loop system,
motivated as follows.  The selected loop system is relatively bright,
well outstanding at the solar limb. We might then suppose relatively
large emission measure, high plasma pressure, and assuming that it is a
steady structure, we can also infer, according to loop scaling laws
(Rosner et al. 1978), high temperature at their apex, maybe higher than
$\sim 2$ MK, which is the temperature obtained from the previous data
analyses.  On the other hand, neither previous analyses include, or
include only partially, a component which instead may be important:
``the loop system is embedded in a hazy background'' (Aschwanden
2001).  This background is significant, especially in the region of the
loop apex, where the loop signal is fainter than elsewhere, due to the
plasma gravitational stratification. One possible hypothesis, that we
pursue here, is that the haze may be the effect of the presence of many
other disordered faint loops, or of a streamer, which extend over the
whole loop system and intersect the analysed structure along the line
of sight. Such background may not be described as a simple hydrostatic
equilibrium, and deserves proper attention, since it may affect
systematically the filter ratio and temperature values. Another item
stimulating our analysis is the finding by Aschwanden (2001) of a
solution with a very small scaling height of the heating and
temperature maxima at the loop footpoints, quite a puzzling result in
the light of previous well-established loop modeling (Rosner et al.
1978, Serio et al. 1981).

Here we revisit the loop system, describing the main loop component
with conventional hydrostatic single loop models and deriving the
background component directly from the data.  We compare the brightness
profiles observed along the loop in the two SXT filter passbands with
and without background subtraction to those obtained from the loop
models. The modeling explores three different scenarios of heating
deposition:  uniform distribution, localized at the footpoints and
localized at the apex, i.e. a more complete sampling than in Aschwanden
(2001).  The model results are compared to data with detailed fitting
procedures, and attention is paid to the values of column depth
obtained to match the observed fluxes.  Section~\ref{sec:model}
describes the analysis of the data and their comparison with loop
models; Sect.~\ref{sec:disc} discusses the results and their
implications.

\section{Data Analysis and Modeling}
\label{sec:model}
\subsection{The Data}

The loop system was observed with Yohkoh/SXT on 3 October 1992.
The data consist of a sequence of full-disk 512$\times$512 pixel images
($\approx 5$" pixel side), basically in the Al.1 and Al/Mg/Mn filter
passbands, taken with various exposure times during the whole day. The
data have been preliminarily processed with the standard Yohkoh data
analysis system (SXT\_PREP).

We have checked that the loop to be analyzed remains quite stable and
stationary during the whole day. In order to enhance the
signal-to-noise ratio, we then average the long-exposure (2.7 s and 15
s) images over the whole day (as done in Aschwanden 2001, 57 images in
the Al.1 filter and 55 images in the Al/Mg/Mn filter).
Fig.~\ref{fig:image} shows the resulting image in the Al.1 filter
passband. We extract the loop flux profiles within the arch-like strip
(14 pixels wide) outlined in the figure. The strip has been divided
into 31 sectors with approximately equal area.  The flux profiles are
then the sequences of Data Numbers averaged over each sector along the
loop. DN fluctuations from each image to the other have been taken as
the main source of data uncertainty.

As remarked by Aschwanden (2001), the loop system is embedded in a
"hazy" background. Here we make the hypothesis that this background is
mostly due to the presence of other disordered faint loops and/or of a
streamer which extend over the whole loop system and intersect the loop
along the line of sight. Since the brightness of the haze is comparable
to the loop brightness in the upper loop region, the effect of the
background may be significant in the analysis and modeling of the loop
emission. Some enhancement of the data quality was achieved by Priest
et al.  (2000) by reducing the radiation scattering due to SXT wings
(Foley 1998), but, as we have checked (using the procedure
SXT\_SCATTER), the amount of scattering ($\le 20$ \%) cannot account
for the high background.

In the hypothesis that the background emission includes disorganized
contributions from other faint structures, our approach has been to
evaluate it with a model-independent procedure and to subtract it from
the emission measured along the loop. To evaluate the background, we
measure the emission in a region below the main visible arcade, where
the emission is relatively faint, unstructured and as far as possible
free of small evident loops.  We then assume that, in the loop region,
the background is stratified in the radial direction from the Sun
center, and extends to involve the whole loop system. We extract the
background fluxes (one for each filter) along the rectilinear strip
(ten pixels wide) outlined in Fig.~\ref{fig:image}, roughly
perpendicular to the disk limb. This strip is chosen to have a signal
decreasing upwards as far as possible smoothly. The resulting radial
background profiles are then subtracted from the respective loop flux
profiles, as a function of the distance of the loop sectors from the
limb. One of the background profiles is shown in
Fig.~\ref{fig:data_models}, described below. The background flux
results to be negligible nowhere along the loop, being of the order of
50\% of the total flux.  This procedure also allows us to limit the
number of free parameters in fitting the dominant loop structure, which
is the main target of the analysis.

We extract the loop flux within the arch-like strip shown in
Fig.~\ref{fig:image} which follows the geometry of the loop system. The
loop enclosed in the strip does not appear to be perfectly
semicircular, and symmetric around an axis perpendicular to the limb.
We have assumed this to be due to a projection effect: the loop may not
lie on a plane perpendicular to the line of sight. Then we have
considered the data along half of the loop (the right leg) and
corrected the coordinates along the loop for the projection effect
(assuming a rotation around the symmetry axis only). As shown by the
modeling results below, this allows us to make detailed comparisons
with semicircular symmetric loop models.

\subsection{Modeling}

As done in Priest et al. (2000) and Aschwanden (2001), we try to compare
the loop data with hydrostatic loop models.
Priest et al. (2000) compare temperature values derived from the data
with those predicted by models.  Temperature values are obtained from
the data with the filter ratio method, i.e. using the ratio of the flux
in two filter passbands. The temperature values are then affected by
two sources of uncertainties (one for each filter Data Number) and by
the presence of high background. Instead of using temperature, our
approach is then to compare flux profiles synthesized from model
results directly to observed flux profiles corrected for background
subtraction. The diagnostics on plasma temperature will then descend
from the parameters of the model (or models) which is (are) found to
provide a good description of the data.

The loop system is stable and steady on a time scale ($\sim 1$ day)
much longer than the relevant physical time scales, i.e. the sound
crossing time and the cooling times ($\la$~few hours). We then consider
standard hydrostatic single loop models as in Serio et al.  (1981),
semicircular, with constant cross-section along the loop and symmetric
with respect to the loop apex. Given the extension of the loop system,
gravity cannot be ignored and the loop is assumed to be perpendicular
to the solar surface. The loop half-length is set to L = 380 Mm as in
Aschwanden (2001). The model is based on the equations of plasma
hydrostatic equilibrium and of energy balance among radiative losses, a
phenomenological term of local heat input in the plasma and thermal
conduction according to Spitzer's formulation. The lower boundary of
the loop is set at $2 \times 10^4$ K, and there we assign the base
pressure and the conduction flux ($F_C \sim 0$).

We generate models with various loop base pressures, corresponding to
loop apex temperature logarithmically spaced (with step $\sim 0.05$) in
the range 1 -- 10 MK, for three different distributions of the heating
along the loop: uniform heating, heating localized at the footpoints,
and heating localized at the loop apex.  The heating at
the footpoints is described with a heating distribution exponentially
decreasing upwards from the loop base, with a scale height
$s_H = 127$ Mm, i.e. 1/3 the loop half length.  With this heating
distribution we explore solutions for loop maximum temperatures below
$\sim 5$ MK. We do not consider models with significantly smaller
heating scale heights: they lead to loops with temperature maximum
localized in the region of the loop footpoints. Such loops have been shown
to be unstable (Rosner et al. 1978, Serio et al. 1981), and therefore
unlikely to describe the loop analyzed here, which appear to be
steady for a time of the order of a whole day.  For the cases of
heating  at the apex, we assume a Gaussian distribution centered at the
apex, with width $\sigma_H = 10$ Mm.

\begin{table*}
\begin{center}
\caption[]{\label{tab:data_models} Parameters of loop models}
\begin{tabular}{lccccccccc}
\hline
&&&&&\multicolumn{2}{c}{Al.1}&\multicolumn{2}{c}{AlMgMn}&\\
N.&Model$^a$&$P_0^b$&$T_{max}^c$&$H_0^d$&Col.Dep.$^e$&$\chi
^2_\nu$&Col.Dep.$^e$&$\chi^2_\nu$&Shift\\
&&dyne cm$^{-2}$&MK&$10^{-3}$ erg cm$^{-3}$ s$^{-1}$&pix&&pix&&pix\\
\hline
\multicolumn{10}{c}{Unsubtracted data - Uniform heating} \\
\hline
1 & L & 0.13 & 2.3 & 0.007 & 820 & 370 & 830 & 240 & 0 \\
2 & B & 0.38 & 3.3 & 0.025 & 17 & 65 & 16 & 31 & 0 \\
3 & H & 1.5 & 5.3 & 0.14 & 0.51 & 260 & 0.43 & 110 & 0 \\
\hline
\multicolumn{10}{c}{Unsubtracted data - Base heating} \\
\hline
4 & L & 0.09 & 1.7 & 0.014 & 3800 & 4300 & 5000 & 1500 & 0 \\
5 & B & 0.19 & 2.2 & 0.033 & 300 & 21 & 310 & 30 & 0 \\
6 & H & 0.53 & 3.1 & 0.12 & 7.0 & 120 & 6.4 & 47 & 0 \\
\hline
\multicolumn{10}{c}{Unsubtracted data - Top heating} \\
\hline
7 & L & 0.07 & 2.2 & 0.09 & 20000 & 4600 & 25000 & 1500 & -6 \\
8 & B & 0.15 & 2.9 & 0.23 & 740 & 230 & 730 & 88 & -6 \\
9 & H & 0.7 & 4.9 & 1.5 & 4.9 & 780 & 4.4 & 340 & -6 \\
\hline
\multicolumn{10}{c}{Bkg-subtracted - Uniform heating} \\
\hline
10 & L & 0.09 & 2.0 & 0.004 & 1300 & 15 & 1600 & 31 & 0 \\
11 & H & 0.38 & 3.3 & 0.025 & 6.5 & 24 & 6.3 & 78 & 0 \\
12 & B & 3.0 & 6.7 & 0.33 & 0.05 & 11 & 0.04 & 19 & 0 \\
\hline
\multicolumn{10}{c}{Bkg-subtracted - Base heating} \\
\hline
13 & L & 0.09 & 1.7 & 0.014 & 1700 & 47 & 2100 & 100 & 0 \\
14 & B & 0.13 & 1.9 & 0.021 & 500 & 5.0 & 570 & 7.8 & 0 \\
15 & H & 0.53 & 3.1 & 0.12 & 2.7 & 13 & 2.6 & 31 & 0 \\
\hline
\multicolumn{10}{c}{Bkg-subtracted - Top heating} \\
\hline
16 & L & 0.07 & 2.2 & 0.09 & 8500 & 53 & 10000 & 130 & -6 \\
\it 17 & \it B & \it 0.30 & \it 3.7 & \it 0.5 & \it 22 & \it 0.50 & \it 20 & \it 1.7 & \it -6 \\
18 & H & 3.0 & 8.3 & 9.6 & 0.09 & 2.8 & 0.07 & 8.1 & -6 \\
\hline
\end{tabular}
\end{center}
\noindent
$^a$ - Label of the model as in Fig.~\ref{fig:chi2}, B, L and H
indicate "Best fit model", "Lower temperature model" and "Higher temperature
model", respectively. \\
\noindent
$^b$ - Pressure at the loop footpoints \\
\noindent
$^c$ - Maximum loop temperature \\
\noindent
$^d$ - Maximum heating rate per unit volume \\
\noindent
$^e$ - Column depth obtained from normalization of the emission
synthesized from the loop models to best-fit the data.
\end{table*}

From the temperature and density profiles obtained from the models we
synthesize the loop emission filtered through the two SXT filter
passbands, with the same procedure as in Aschwanden (2001).  We then
perform a best-fit procedure to the data along half of the loop, by
finding analytically the normalization factor which minimizes the
$\chi^2$ of each model to the data points. This normalization factor is
the only parameter of the fitting and represents the plasma column
depth of the loop system along the line of sight.  We then look for the
models yielding the lowest $\chi^2$ values among the models with
different maximum temperatures and the same heating distribution, and
then for the best-fit model among those with different heating
distributions.  This procedure is repeated on the data, both with and without
background subtraction.  In order to make the fitting in some way
independent of possible errors in the determination of the loop length
and of the real position of the relevant footpoint (see also
the problem of submerged footpoints
presented in McKay et al. 2000), the first three
data points from the left are excluded from the $\chi^2$ calculation,
and we repeat
the fitting for two different values of the relative shift of the model
to the data, one assuming the model loop footpoint coincident with the
data origin, and the other shifted backwards by $2 \times 10^9$ cm. 
In the following we will show results only for the value of the shift
yielding the better $\chi^2$ value. We cannot exclude also some effects
on the results due to loop opening at the base the loop.
These are not expected to be important in corona (Serio et al. 1981),
and may somewhat influence the normalization factor (Aschwanden 2001), but
warnings on this have been raised due to the possibility that the coronal
footpoint of a loop may have any base temperature since it may virtually be
thermally isolated from the chromosphere (Dowdy et al. 1985).

Table~\ref{tab:data_models} and Figs.~\ref{fig:data_models} and
~\ref{fig:chi2} show some relevant results of the modeling, separating
the cases of data without (hereafter {\it unsubtracted data}) and with
background subtraction, and of models with uniform, base and top
heating.  Table~\ref{tab:data_models} shows the significant parameters
of the models in Figs.~\ref{fig:data_models} and
~\ref{fig:chi2}, and in particular
the pressure at the base of the model loop, its maximum temperature (at
the top of the model loop), the (maximum) heating rate per unit
volume, the loop plasma column depth derived as 
normalization factor, the reduced $\chi^2$ values (11 d.o.f.),
respectively obtained for the two filters, and the shift value yielding
the best $\chi^2$.

In Fig.~\ref{fig:data_models}, each panel shows the data
(data points) in both filters, the best-fitting
model (solid lines) and other two models, one with lower and one with
higher maximum temperature, for comparison. The origin of the reference
system is at the loop footpoint. The panels on the left show data
without background subtraction, those on the right data with background
subtraction. From top to bottom, Fig.~\ref{fig:data_models} shows
fitting results for models with uniform heating, heating at the
footpoints and heating at the apex, respectively.  The upper left panel
also shows one of the background profiles which are subtracted from the
data to obtain the values shown in the lower two panels. The (similar)
profile in the other filter passband is not shown for the sake of
clarity.

Fig.~\ref{fig:chi2} shows plots of the minimum $\chi^2$ obtained from
fitting the data with models with different maximum temperature for the
two filters (data and models order as in Fig.~\ref{fig:data_models}).

Several considerations are in order. The trends and values of the
unsubtracted data are similar to those shown in Aschwanden (2001). Due
to their lower statistics, the error bars of the background-subtracted
data are larger than those of the unsubtracted data, and they are
largest close to the loop apex, and in the Al/Mg/Mn filter passband.
The background profile is monotonic, decreases toward the loop apex,
and its trend is only partially similar to that of the full loop signal
along the loop. The background-subtracted flux becomes quite small
around the loop apex, below 1 DN/s in the Al/Mg/Mn filter passband.

Loop models with uniform heating resemble the trends of the
unsubtracted data more closely than those of the background-subtracted
data, although the latter ones yield better $\chi^2$ values (because of
the lower data statistics). The loop model with uniform heating
best-fitting both the unsubtracted data and the background-subtracted
data is the one with maximum temperature 3.3 MK (model 2 and 11 in
Table~\ref{tab:data_models}) and a pressure $\sim 0.4$ dyne cm$^{-2}$.
The values of column depth obtained in the two filters are similar and
reasonable ($\la 20$ pixels), but the fitting is poor (high $\chi^2$)
in the region around the loop apex (Fig.~\ref{fig:data_models}).
Notice that the 2 MK models (e.g. model 1 and 10 in
Table~\ref{tab:data_models}) yield an even worse fitting in the same
region, and too large column depths, in agreement with Aschwanden
(2001).

Models with heating at the footpoints are more in agreement with data
both with and without background subtraction. Unsubtracted data are best-fit by
the model with temperature 2.2 MK at the top (base pressure 0.2 dyne
cm$^{-2}$, model 5 in Table~\ref{tab:data_models}). However, the
$\chi^2$ value is high, and the column depth is unreasonably large
(larger than the solar radius). Background-subtracted data are best-fitted
by a model with a lower maximum temperature (1.9 MK, model 14 in
Table~\ref{tab:data_models}), which yields a
better (but still not statistically acceptable) $\chi^2$ value, but an
even larger column depth. Notice that both best-fit base-heated models,
as well as those with uniform heating, do not describe well the flux at
the footpoint, especially in the softer Al.1 filter passband, and
systematically overestimate the flux at the loop apex.

Going to models with heating at the loop top, the one best-fitting the
unsubtracted data (2.9 MK, model 8 in Table~\ref{tab:data_models}) largely
underestimates the emission in the apex region, and yields very high
$\chi^2$ values and very large column depth values. On the other hand,
the background-subtracted data appear to be adequately described in
both filters by the (best-fit) top-heated loop model with a maximum
temperature of 3.7 MK (model 17 in Table~\ref{tab:data_models}): the
observed flux is well described anywhere along the loop, at the
footpoint and at the apex altogether; the values $\chi^2_\nu \sim 1$
make the fitting statistically acceptable in both filters, and make
such model {\it the absolute best-fitting one among all the models
explored here}. Furthermore, the column depth values ($\la 10^{10}$ cm)
are reasonable for large scale or arcade structures.  The low emission
obtained at the apex with the best-fit top-heated model is the natural
result of the localized heating excess, which steepens the temperature
distribution to be more peaked at the apex, and makes the density
decrease to maintain the pressure balance
(Fig.~\ref{fig:bestmodel}). The maximum temperature and the pressure at
the footpoint (0.3 dyne cm$^{-2}$) are not unreasonably high for large
scale loops.

The fitting of the unsubtracted data invariably shows
well-defined $\chi^2$ minima (Fig.~\ref{fig:chi2}, left column). This
is true also for background-subtracted data, with the exception of the
fitting with models with uniform heating, which shows quite a constant
$\chi^2$ with model maximum temperatures, and similar $\chi^2$ values
at low ($\sim 2$ MK) and high ($> 5$ MK) temperatures (but with very
different column depth values). High temperature top-heated models
fitting the background-subtracted data is the only combination to yield
statistically acceptable results.

Notice in Fig.~\ref{fig:chi2} that fitting unsubtracted data 
in the Al/Mg/Mn filter passband yields generally better
results than in the Al.1 filter passband.  The opposite is true for
background-subtracted data.

\section{Discussion and Conclusions}
\label{sec:disc}

This work has been stimulated by the contrasting results obtained from
the analysis of the same loop system observed with Yohkoh/SXT by two
subsequent works. In the earlier one (Priest et al. 2000) the data, and
in particular the temperature distribution along the loop, were best
matched by a loop model with a uniform heating distribution. The later
one (Aschwanden 2001) found instead that a model dominated by a heating
deposited at the loop footpoints, with a minor cooler component,
provides more self-consistent results, because it predicts also
reasonable values of the plasma column depth.

The present work extends the analysis of the same loop system, by
including models with heating at the loop apex, and by considering a
significant background emission possibly due to the intersection with
other structures, such as disorganized fainter loops and/or a streamer,
along the line of sight, which adds to the loop emission. The analysis
shows that, if background emission is not subtracted, none of single
loop models explored acceptably fits the data, indicating that the
system cannot be described only as in a simple hydrostatic
equilibrium.  A base-heated loop model with maximum temperature 2.2 MK
yields the best results on unsubtracted data, but, besides the high
$\chi^2$, it involves a column depth larger than the solar radius and
trends not matching the data in some regions along the loop.  We did
not explore loop models with heating confined in a narrow region near
the loop footpoints, as in Aschwanden (2001): such heating leads to a
temperature maximum localized at the loop footpoints, and such
configuration is well-known to be unstable (Rosner et al. 1978, Serio
et al. 1981).  On the other hand, the coronal plasma density of
$10^{10.3}$ cm$^{-3}$ predicted by the base-heated ``best'' model loop
in Aschwanden (2001) is unusually large for large scale coronal
structures (e.g. Vaiana \& Rosner 1978).  The best-fit uniformly-heated
loop (with maximum temperature 3.3 MK) matches the data slightly worse
than the best-fit base-heated model, but implies much more reasonable
values of the column depth.

We do find a model fitting acceptably the background-subtracted data:
a loop heated at the loop apex, with a maximum temperature of 3.7 MK
(model 17 in Table~\ref{tab:data_models}).  The corresponding column
depth ($\sim 1/10$ of the loop length) represents a reasonable aspect
for a loop system.  Although the lower statistics of
background-subtracted data should allow in general to obtain more
easily an acceptable fitting, the model above is {\it the only one} to
fit the data with high statistical significance. The combination of an
acceptable fitting and of a realistic column depth provides a scenario
globally self-consistent and consistent with the data.

Several considerations are now in order:

\begin{itemize}

\item 
Our best model is a loop heated at the apex, with a relatively high
(still reasonable) temperature, a pressure of 0.3 dyne cm$^{-2}$ at the
footpoints and a density decreasing to $\sim 3 \times 10^7$ cm$^{-3}$
close to the apex. Such pressure and density values are not unusual for
large scale structures (e.g. Vaiana \& Rosner 1978). A loop heating
localized in corona is not a new result (e.g. Reale et al. 2000).

\item The best fitting results have been obtained after subtraction of
an unstructured background emission, extrapolated directly from the
data, in the hypothesis that it is due mostly to chance alignment of
disordered faint loops and/or of a streamer along the line of sight,
and therefore not to be modelled in hydrostatic equilibrium.  This is
one possible description of the scenario, not necessarily the best one,
but, in our opinion, it is at least as realistic as other more local or
model-dependent estimations of the background.

\item High data statistics, obtained in this case by averaging over
tens of exposures, is important to discriminate among loop models with
different temperature and heating location.  Fig.~\ref{fig:data_models}
shows that larger error bars would made such discrimination much
harder. This shows, once again, that the task of fitting data with loop
models is not at all trivial, and requires both detailed modeling and
high quality data.

\end{itemize}

Of course, this work suffers from limitations. The background
evaluation may not be unique, and a single loop model may be a
simplified approximation to describe part of an arcade.  Also the
exploration of the loop parameter space is forcedly limited, and we
cannot exclude that a more refined tuning of the free parameters, and
including other effects, such as strong geometrical variations at the
base of the loop, could bring to equally good or better fitting
results.  However, our analysis and modeling lead to a statistically
acceptable description of the data, and to physically sound results
(loop heating and plasma conditions, column depth), and appears
therefore to be adequate to the data quality.

This work shows that the analyzed observational data can be described
with a conventional hydrostatic and non-isothermal loop models and with
an unstructured vertically stratified spurious emission.  The
deposition of heat at the apex of large scale structures is a different
result from those of both previous works, and may not be in agreement
with recent results based on the analysis of data from the TRACE
mission (Aschwanden et al. 2001).  Since there are serious independent
arguments which make the analysis and temperature diagnostics with
narrow band EUV instruments a very delicate issue (Schmelz et al.
2001), we believe that, at variance with Aschwanden (2001), the
problems of the diagnostics of the heating function in corona and of
the coherent interpretation of observations with different telescopes
(e.g. Yohkoh and TRACE) are both still open.

We do not pretend to put a conclusive word on this topic, nor claim
that our results are necessarily the best ones, rather we want to
stimulate the community toward a more careful attention to aspects of
the data analysis, such as the evaluation of background emission,
important for a correct interpretation and modeling of the data. Other
authors (McKay et al. 2000) come to similar conclusions, although with a
totally different analysis. The present work also emphasizes the
importance of the selection of the loops to be analyzed: loops embedded
in crowded and complex coronal regions may not be the best ones,
because emission from structures of different kind and characteristics
may intersect along the line of sight, which may be difficult to
evaluate. Further extensive and systematic analysis and modeling of
bright, and as far as possible isolated, loop structures may provide
more constraints on the issue of coronal heating deposition.

\bigskip
\bigskip
\acknowledgements{I would like to thank Salvatore Orlando, Paola Testa
and Salvatore Serio for their helpful criticism and interest in this
work.  This work was supported in part by Agenzia Spaziale Italiana and
by Ministero dell'Istruzione, dell'Universit\`a e della Ricerca. }

{}

\newpage

\figcaption[fig1]{Region of the solar corona including the analyzed
loop system observed with Yohkoh/SXT in the Al.1 filter passband. The
data to be compared with models have been extracted within the marked
arch-like strip ({\it grey}), and averaged in each sector inside the
strip.  The background emission has been evaluated in the other ({\it
black}) strip.  \label{fig:image}}

\figcaption[fig2]{Results of fitting the data along
half of the loop system with loop models: observed flux in the Al.1 filter passband
({\it triangles}) and in the Al/Mg/Mn filter ({\it diamonds}), and the
respective profiles from the best-fit model ({\it solid lines},
labeled B in Fig.~\ref{fig:chi2}) and from two other models, one with
lower ({\it dotted}, L) and one higher ({\it dashed}, H) maximum
temperature, for comparison.  The origin of the reference system is at
the loop footpoint.  The upper left panel also shows the background
profile in the Al/Mg/Mn filter ({\it stars}).  Unsubtracted data 
are shown on the {\it left}, background-subtracted data on
the {\it right}.
{\it From top to bottom}:  
fitting with the models with uniform heating, heating at the footpoints, and
heating at the apex.
\label{fig:data_models}}

\figcaption[fig3]{Results of the fitting of loop models with data along
half of the loop system:
$\chi^2$ obtained with the models with different maximum
temperature, for the Al.1 filter ({\it symbols}) and for the Al/Mg/Mn
filter ({\it dashed line}). Order as in Fig.~\ref{fig:data_models}.
\label{fig:chi2}}

\figcaption[fig4]{Temperature and density distributions along half of
the loop of the best fit loop model with heating at the top (model 17
in Table~\ref{tab:data_models}, {\it solid lines}) and of other two
models, one with uniform heating (model 2 in
Table~\ref{tab:data_models}, {\it dashed lines}), and one with heating
at the footpoints (model 5 in Table~\ref{tab:data_models}, {\it dotted
lines}). \label{fig:bestmodel}}

\begin{figure}
\epsscale{0.8}
\figurenum{1}
\plotone{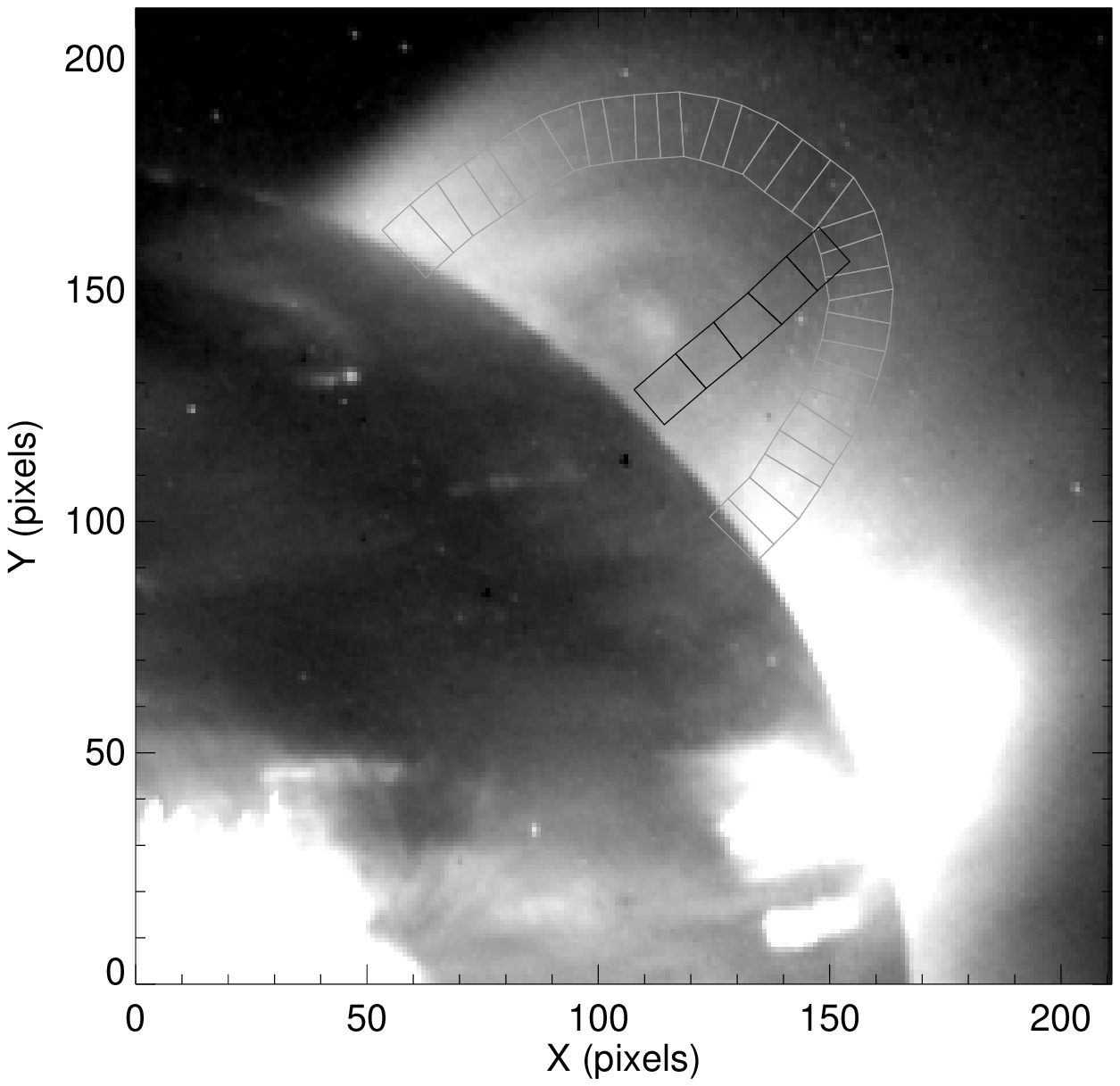}
\caption{}
\end{figure}

\begin{figure}
\epsscale{0.7}
\figurenum{2}
\plotone{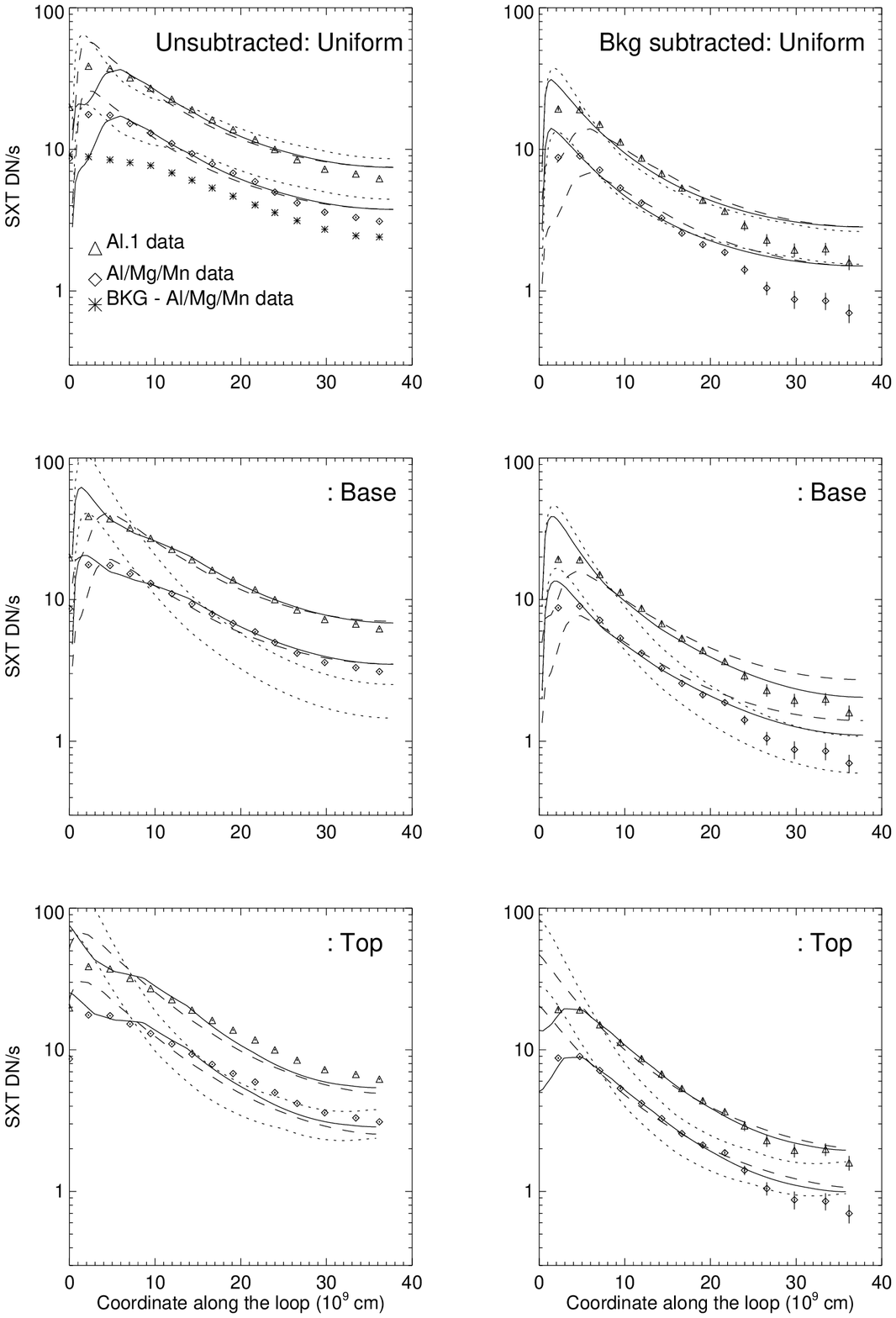}
\caption{}
\end{figure}

\begin{figure}
\epsscale{0.7}
\figurenum{3}
\plotone{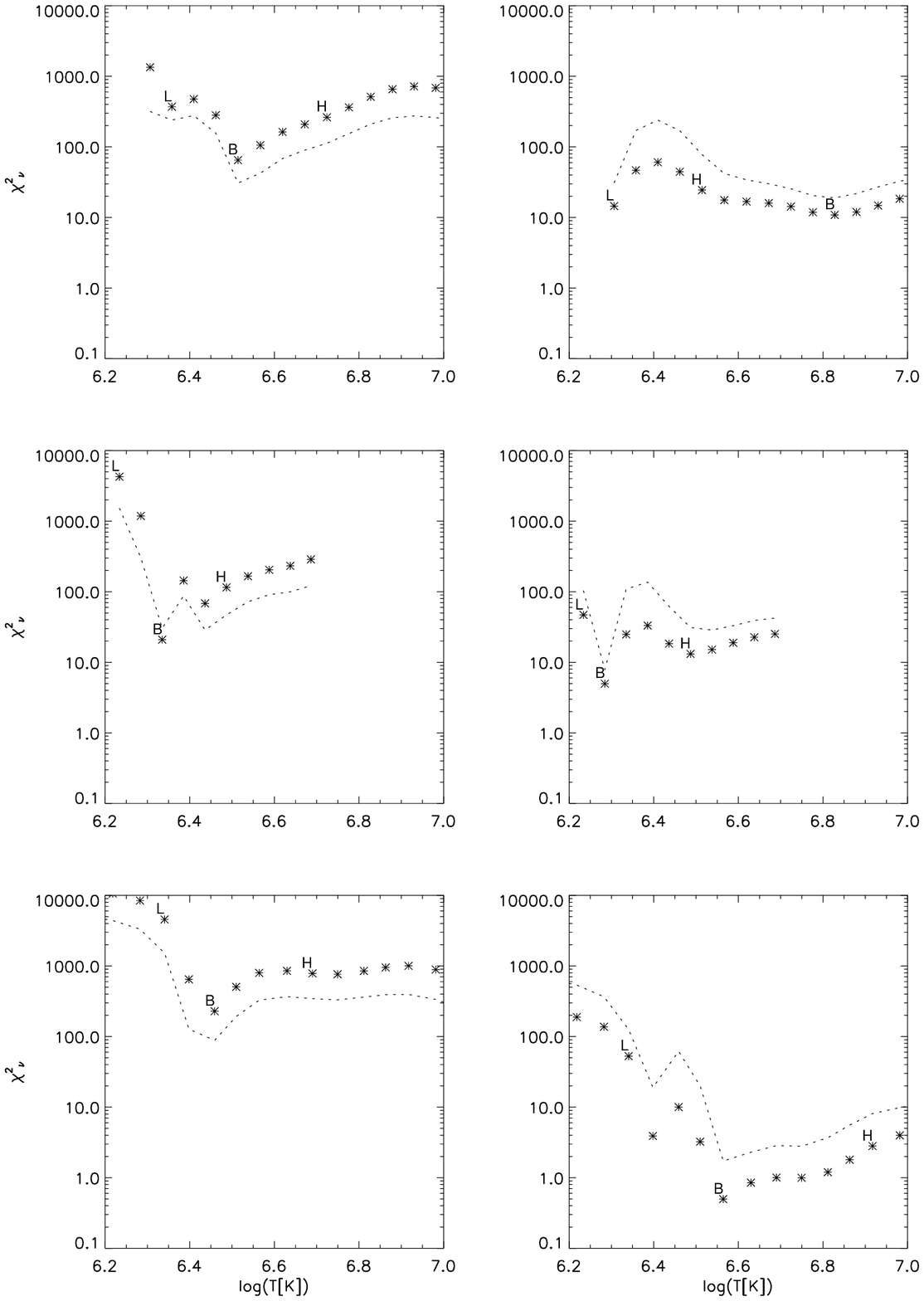}
\caption{}
\end{figure}

\begin{figure}
\epsscale{0.6}
\figurenum{4}
\plotone{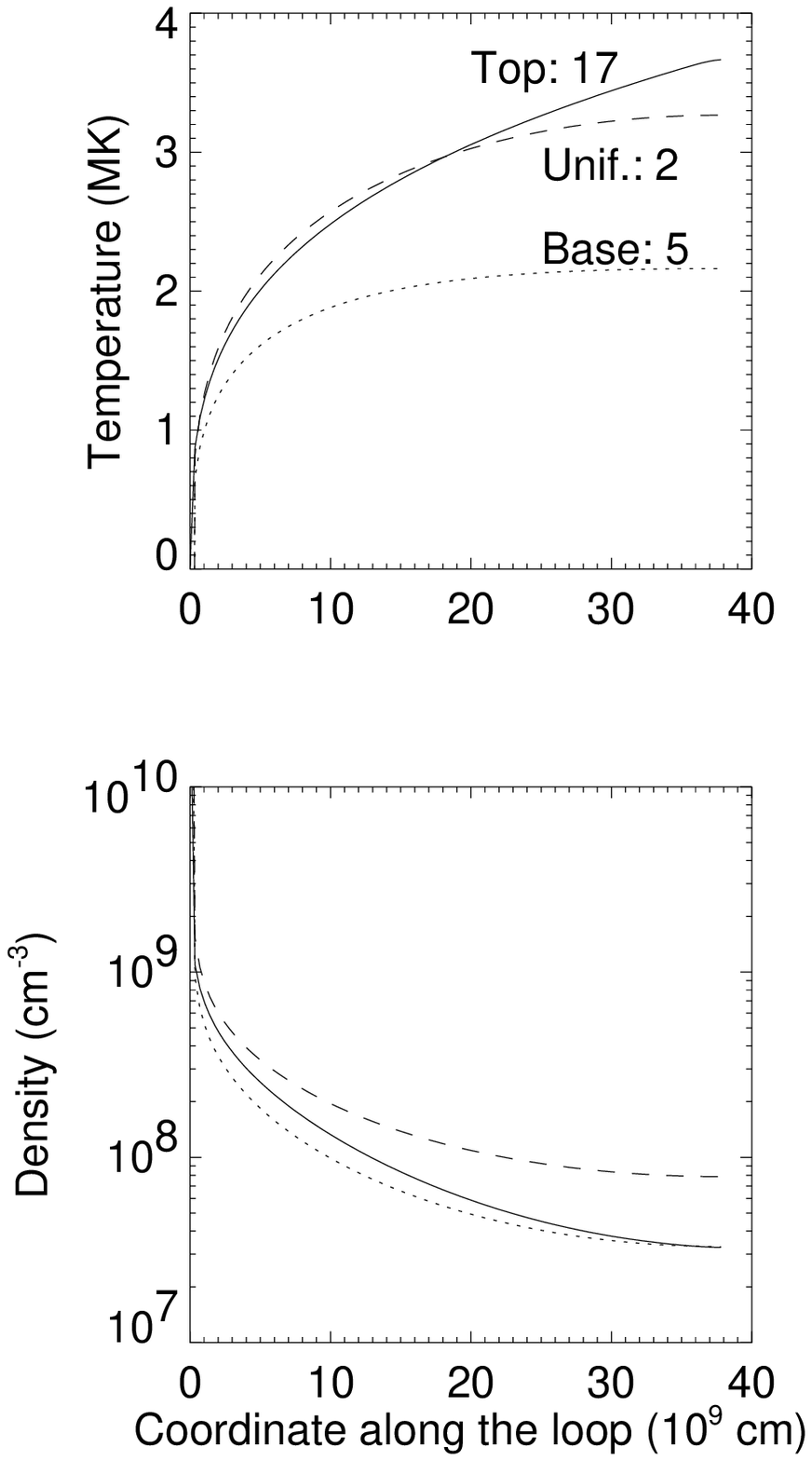}
\caption{}
\end{figure}

\end{document}